\documentclass[useAMS,usenatbib]{mn2e}
\usepackage{longtable}
\usepackage{subfigure}
\usepackage{graphicx}
\usepackage{graphics}
\usepackage{keyval}
\usepackage{trig}

\title[3D numerical model of M17]{3D numerical model of the Omega
Nebula (M17): simulated thermal X-ray emission} \author[Reyes-Iturbide
et al.]{J. Reyes-Iturbide$^{1}$\thanks{E-mail:
jreyes@astrocu.unam.mx (JRI), pablo@nucleares.unam.mx (PFV), 
margarit@astroscu.unam.mx (MR), ary@nucleares.unam.mx (ARG), 
rf.gonzalez@astrosmo.unam.mx (RFG),  esquivel@nucleares.unam.mx (AE)},
P. F. Vel\'azquez$^{2}$, M. Rosado$^{1}$,
A. Rodr\'\i guez-Gonz\'alez$^{2}$,
\newauthor R. F. Gonz\'alez$^{3}$, and
A. Esquivel$^{2}$ \\ $^{1}$Instituto de Astronom\'\i a,
Universidad Nacional Aut\'onoma de M\'exico, Apdo. Postal 70-264,
04510 M\'exico, D. F., M\'exico\\ $^{2}$Instituto de Ciencias
Nucleares, Universidad Nacional Aut\'onoma de M\'exico, Apdo. Postal
70-543, D.F.  M\'exico\\ $^{3}$Centro de Radioastronom\'\i a y
Astrof\'\i sica Te\'orica, Universidad Nacional Aut\'onoma de
M\'exico, Apdo. Postal 3-72 (Xangari), \\ C.P. 58089, 
Morelia, Michoac\'an, M\'exico}
\begin{document}

\date{Accepted ---. Received ----; in original ---}

\pagerange{\pageref{firstpage}--\pageref{lastpage}} \pubyear{2002}

\maketitle

\label{firstpage}

\begin{abstract}

We present 3D hydrodynamical simulations of the superbubble M17, also
known as the Omega nebula, carried out with the adaptive grid code
{\sc yguaz\'u-a}, which includes radiative cooling.  The superbubble is
modelled considering the winds of 11 individual stars from the open
cluster inside the nebula (NGC 6618), for which there are estimates of the
mass loss rates and terminal velocities based on their spectral types.
 These stars are located inside a dense interstellar medium, and they
  are bounded by two dense molecular clouds. 
 We carried out three numerical models of this scenario,
considering different line of sight positions of the stars
(the position in the plane of the sky is known, thus fixed).
Synthetic thermal X-ray emission maps are calculated from the
numerical models and compared with ROSAT
observations of this astrophysical object. Our models reproduce
successfully both the observed X-ray morphology and the total X-ray
luminosity, without taking into account
thermal conduction effects.

\end{abstract}


\section{Introduction}

The mechanical luminosity of a stellar wind ($ \frac {1} {2}\dot M_w
V_{\infty}^2$; where $\dot M_w$ is the mass loss rate and $V_{\infty}$
is the terminal speed of the wind) is typically less than 1 per cent of the
stellar radiative luminosity. The fast winds of massive stars have,
however, a great influence to their surrounding interstellar medium
(ISM). They sweep up the ISM creating a variety of structures, from
small bubbles around single stars to large  
superbubbles around OB associations. The interior of these bubbles or
superbubbles contains shock-heated gas (with temperatures in excess of
$10^{6}$ K), thus emits strongly in X-rays, while the outer shell is
cooler and bright in optical emission lines (Weaver et al. 1977; McCray \&
Kafatos, 1987; Mac Low \& McCray, 1988; Chu et al. 1995).

In the last decades, observations from old X-ray satellites such as
the ROSAT, and newer ones, such as the Chandra and XMM-Newton
observatories, have found diffuse X-ray emission inside OB massive
star clusters.  The diffuse X-ray region is associated to hot gas
contained into a superbubble produced by the interaction of the
individual OB stellar winds with the surrounding environment.
Nevertheless, extremely massive stars from the cluster would rapidly
evolve into supernovae (SNe) explosions, in which case the superbubble will
be produced by the combined action of stellar winds and supernovae. 
In order to isolate the X-ray emission from stellar winds alone
we can study very young superbubbles, where SNe explosions are still not
taking place.

The superbubble M17 contains the massive stellar cluster NGC 6618,
with more than 800 stellar sources (Broos et al. 2007). This cluster
has 100 stars earlier than B9 (Lada et al. 1991) and an average age of
$\sim 1$ Myr (Hanson et al. 1997). With such a short age it is
unlikely to have produced any SN explosion; in a cluster with 100 O
type stars the first SN explosion would occur after $\sim$ 4 Myr
(Kn\"odlseder et al. 2002). Therefore, M17 provides an ideal
laboratory to observe and study the diffuse X-ray emission as produced
by stellar wind collisions alone. This kind of superbubbles are
called `quiescent superbubbles'.

This superbubble is highly asymmetric due to the interaction
with the edge of a massive molecular cloud, M17SW, located to the West
of M17 (Lada et al. 1974). In this direction the bubble encounters
resistance, while it can expand more freely to the East.  This
asymmetry is evident in a large-scale map of M17 at $\lambda$ 21 cm
(Felli et al. 1984), where two intersecting clouds or  `bars' are
observed in emission.  These have been called the northern (N) and
southern (S) bars. The projected size of each bar is of $\sim$ 5.7 pc,
forming an angle of $\sim 45\degr$ with each other on the plane of the
sky (Felli et al. 1984; Brogan \& Troland 2001).

\begin{figure}
\includegraphics[angle=0,width=\columnwidth]{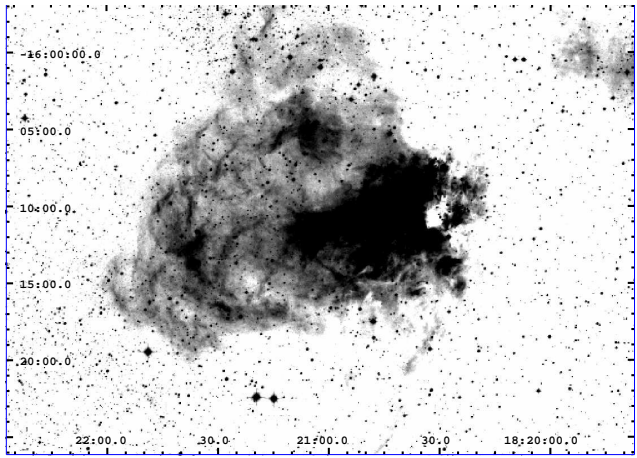}

  \caption{Image of the omega nebulae from the Digital Sky Survey
    (DSS). It is illuminated by the massive stellar cluster NGC 6618,
    the centre of which is located behind substantial obscuring
    material, with $A_{\nu}\sim 8$ mag. }
\label{f0}
\end{figure}

The centre of the stellar cluster NGC 6618 consists of a `ring' of
seven O-type stars (Hanson et al. 1997) which are located between the two bars.
These stars must be the dominant sources of ionising photons on 
the nebula, in fact  diffuse X-ray emission is observed to be well confined
within the external shell of the optical nebula (Dunne et al. 2003). Figure 
\ref{f0} shown a DSS image of this object. The ring of seven O-type stars
is located on the West, where is substantial obscuring material, with
$A_{\nu}\sim 8$ mag (Hanson et al. 1997). 

The stellar winds of these seven stars strongly contribute
to the formation of a common `stellar cluster wind' \citep{canto00}.  
The collision of these winds with the surrounding inhomogeneous ISM
produces the diffuse X-ray emission observed by ROSAT and Chandra
satellites.

In ROSAT images, M17 is observed as a blister-like structure, with an
overall diameter of $\sim 10-12~\mathrm{pc}$, if a distance of
$1.6~\mathrm{kpc}$ is considered (Nielbock et al. 2001). The emission
peak is located inside the stellar cluster.  This suggested that the
cluster is the origin of the diffuse X-ray emission, and was later
confirmed by Chandra observations (Townsley et al. 2003), where the angular
resolution allowed to see the individual sources. M17
was observed with ROSAT Position Sensitive Proportional counter (PSPC)
for $6.7$ ks in the $0.1-2.4$ keV band (Dunne et al. 2003). They
found a total diffuse X-ray luminosity of $\sim 2.5 \times 10^{33}
\mathrm{erg~s^{-1}}$ with kT = 0.66-0.78 keV
($T = 7.7-9.1 \times 10^{6}$ K) and $N_H = (1-5)\times
10^{21}~\mathrm{cm^{-2}}$.

Several models have tried to reproduce the total X-ray luminosity
$\mathrm{L_X}$ from superbubbles, often taking into account the effects of the
electron thermal conduction. However, the $\mathrm{L_X}$ reproduced by models
where thermal conduction is included (Weaver et al. 1977) is usually two or
three orders of magnitude higher than the observed values, while models where
thermal conduction is ignored predict a $\mathrm{L_X}$ that is typically two
orders of magnitude lower than observed.  Most of the time the descriptions
are rather simple, they consider for instance that the cluster wind interacts
with an homogeneous and isotropic ISM, and neglect the effects of radiative
cooling. For M17 this approach is not adequate because it is a star forming
region surrounded by asymmetrically distributed high density molecular clouds.

In this work we present three-dimensional numerical simulations
based on the properties of M17. The simulations were carried out with the
adaptive grid code {\sc yguaz\'u-a}, and considered the main characteristics of
the ISM that surrounds M17.  Synthetic thermal X-ray emission maps were
obtained from the numerical results in order to compare the
morphology and the total luminosity directly with ROSAT X-ray observations
(with a field of view of $\sim 2\degr$). 

This manuscript is organised in the following way: in section 2 we
explain the modelling of M17, the initial conditions of numerical
simulations, and describes the simulation of the thermal X-ray
emission; an archival ROSAT image of this superbubble is presented in
section 3; the results and comparison with observations are given in
section 4; and finally in section 5 we summarise our conclusions.

\begin{table*}
 \centering
 \begin{minipage}{140mm}
  \caption{OB Stars in M17 and their stellar wind parameters.}
  \begin{tabular}{@{}lccccccc@{}}
  \hline
   Name & R.A.(J2000.0)& DEC.(J2000.0)&  Optical & K-Band  & $V_{\infty}$ & $ \dot{M}$&\\
&(\fh~\fm~\fs) & (\degr~\arcmin~\arcsec) &  Spectral type & Spectral type
   & ( $\mathrm{km~s^{-1}}$ )& ($\mathrm{M_{\odot}~yr ^{-1}}$ ) & References \\ 
\hline
B98  & 18 20 35.45 & $-$16 10 48.9 & O9 V&kO9-B1& 2200 &$2.6\times10^{-7}$ & 1,2,3 \\
B111 & 18 20 34.55 & $-$16 10 12.1 & O5 V&kO3-O4& 3250& $1.86\times 10^{-6}$ & 3  \\
B137 & 18 20 33.14 & $-$16 11 21.6& &kO3$-$O4&3370 & $2.4\times 10^{-6}$ &4,5 \\
B164 & 18 20 30.92 & $-$16 10 08.0 & O7$-$O8 V&kO7-O8 &3015 &$2\times 10^{-6}$&1,4\\
B174 & 18 20 30.54&$-$16 10 53.3 & &k03$-$O6 &3370& $6.5\times 10^{-7}$&1,4 \\
B181 & 18 20 30.30 &$-$16 10 35.2& &kO9$-$B2&2200 & $2.6 \times 10^{-7}$&1,2,3 \\
B189 & 18 20 29.92 &$-$16 10 45.5 &O5 V& kO3$-$O4&3250 & $1.86\times 10^{-7}$&3 \\
B260 & 18 20 25.94 &$-$16 08 32.3 & O7$-$O8 V&kO3-O4&3015 &$2\times 10^{-6}$ &1,4 \\
B289 & 18 20 24.45 &$-$16 08 43.3 & O9.5 V&& 1500&$2.5 \times 10^{-7}$&1,6 \\
B311 & 18 20 22.76 & $-$16 08 34.3& & kO9$-$B2 &2200 &$2.6 \times 10^{-7}$&1,2,3  \\
OI 345 & 18 20 27.52 &$-$16 13 31.8  & O6 V &kO5-O6&3065 & $6.5\times 10^{-7}$&1,4\\

\hline

\label{tabstar}
\end{tabular}

{References.--(1) de Jager et al. 1988; (2) Wilson \& Dopita 1985; (3)
  Leitherer 1988; (4) Prinja et al. 1990; (5) Lamers \& Leitherer 1993; (6)
  Fullerton et al. 2006}

\end{minipage}
\end{table*}

\section{Modelling the M17 nebula}

As it was mentioned above, the stars that belong to the NGC 6618
cluster are too young to generate SN explosions. For this reason, the
M17 nebula was modelled by the interaction of stellar winds alone.
We only considered the stars that dominate the mechanical luminosity of the
cluster, they have high terminal velocities and mass loss rates, the
values are based on the spectral types reported by \citet{hanson}, we
list them in Table \ref{tabstar}. The values for the terminal velocity
and mass loss rate correspond to the upper limits reported in the literature
(de Jager et al. 1988; Wilson \& Dopita 1985; Leitherer 1988; Prinja
et al. 1990; Lamers \& Leitherer 1993; Fullerton et al. 2006). This
has been done with the intention to increase the X-ray luminosities
obtained from our simulations, and to compensate for the fact that
all the other wind sources have been disregarded. Furthermore, a recent
study of \citet{hoff08} has shown that the number of stars in the NGC 6618
cluster is even higher than previous estimates.

For the 3D numerical simulations we consider that the
plane of the sky corresponds to the $xy-$plane of our simulation.
Table \ref{tabstar} gives the position of the stars in equatorial
coordinates (J2000), which can be translated to parsecs considering
that the cluster is at a distance of $1.6~$kpc.

\begin{figure}
\includegraphics[angle=0,width=\columnwidth]{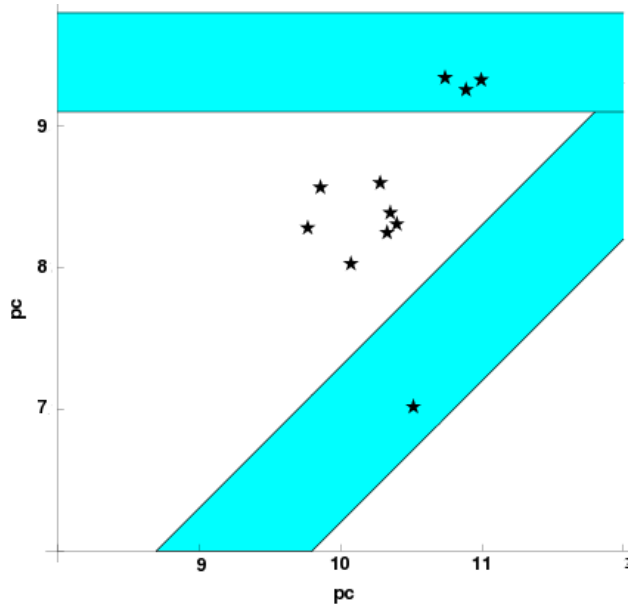}

  \caption{Stellar positions in the plane sky (the $xy-$~plane). These
    positions were calculated considering the R.A and Dec. positions listed on
    Table \ref{tabstar} and a cluster distance of 1.6kpc. The shaded area
    represents the location of the dense molecular clouds that surround the M
    NGC 6618 cluster.}
\label{f1}
\end{figure}

\begin{figure}
\includegraphics[angle=0,width=\columnwidth]{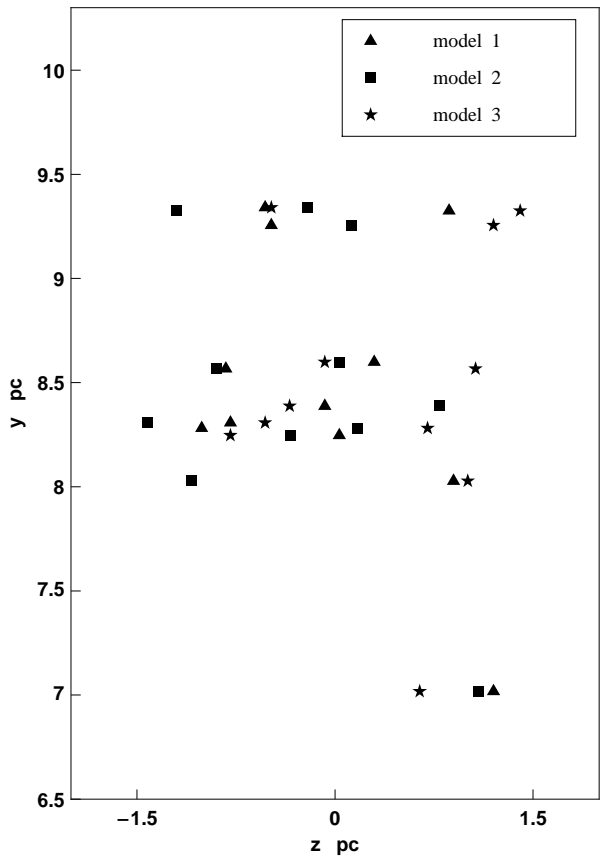}
\caption{Stellar configuration in the $yz-$plane for each of the models,
  as indicated in the legend.}
\label{f2}
\end{figure}

Since we do not know the individual line-of-sight distance
($z-$coordinate) to the stars, we produced three different
realisations of randomly picked positions in $z$, while keeping the
same $xy$ configuration  (see Figure~\ref{f1}). The maximum of the
distribution from which the $z$ positions were sampled was set to the
maximum separation in the plane of the sky, and the mean was set to
the distance to the cluster ($1.6~\mathrm{kpc}$.)
The resulting $yz$ distributions are shown in Figure~\ref{f2}.

The NGC 6618 cluster is bounded by two molecular clouds at the North and
South-West, observed in HI (Felli et al. 1984).  These clouds have the
appearance of a wedge which confine the resulting stellar cluster wind, and
produce an elongated structure to the East.  We model the clouds as two
bars one of them is horizontal (we will reefer to it as the `top bar')
and the other one tilted 45\degr in the $xy$-plane (which we will call
the `bottom bar'). Their surfaces are flat, and
cover the entire extent of the computational domain along the $z$~axis
(i.e. the geometry displayed in Figure \ref{f1} is the same for all
values of $z$).
The bars have a high density contrast with respect to the surrounding ISM
(of two orders of magnitude).  The stars (wind-sources) were placed inside the
wedge formed by the two bars, as shown schematically in Figure~\ref{f1}.

\subsection{Initial setup}

We have carried out the 3D numerical simulations with the {\sc
  yguaz\'u-a} hydrodynamical code \citep{raga00,raga02}.  The code
integrates the gasdynamic equations with a second-order accurate
implementation of the flux vector splitting method of van Leer (1982)
on a binary adaptive grid.  
This code was first tested comparing simulations of shock waves
and laser laboratory experiments (one bubble shock wave, Raga et
al. 2000; two interacting bubbles, Vel\'azquez et al. 2001). In last 8
years, the {\sc yguaz\'u-a} code has been widely used to model several
different astrophysical scenarios, such as Herbig Haro objects
\citep{raga02,masciadri02}, supernovae remnants
\citep{schneiter06,reyes08}, jets in planetary nebulae
\citep{velazquez04,guerrero08}, interacting winds \citep{ricardo04,ary07},
and exoplanets \citep{schneiter07}.

For this problem we use a computational domain with a physical size of
($12\times 12\times 4$) pc ($x-$,$y-$, and $z-$axis, respectively),
and allow five levels of refinement, such that the resolution at the
finest level of is  $7.27\times 10^{16}$~cm. This would
correspond to  $512\times 512 \times 128$  pixels in a uniform grid.
We chose this spatial resolution ($3\arcsec$ if a distance of
$.1.6~\mathrm{kpc}$ to M17 is assumed) to match the resolution 
of ROSAT raw observations ($\sim 4\arcsec$, see \S 3.)

The stellar winds are imposed 
in spheres (centred at the stellar positions) of radius $R_w=4.4 \times 
10^{17}$~cm, which corresponds to 6 pixels at the maximum resolution of the 
adaptive grid (always present at the wind sources).
Within these spheres, we inject at every time-step material at $T_w=1000$~K, 
with the outward velocity given in Table~\ref{tabstar} for each star. 
The density inside the spheres follows an $r^{-2}$ law (where $r$ is 
the radial coordinate measured outwards from the stellar position),
scaled to yield the mass loss rates in Table~\ref{tabstar}.

The stars and the bars are located in the top-right quadrant of
the numerical domain, in the $xy-$ plane. The bars were
initialised with a uniform number density of 1000 cm$^{-3}$ and a
temperature of 400 K in order to stay in pressure equilibrium with an
ISM that has a density of 10 cm$^{-3}$, and a temperature of 10$^4$~K.

\subsection{Simulated X-ray emission}

The numerical simulation provides us with density and
temperature distributions for each model, which we combine with
synthetic X-ray spectra to simulate the X-ray emission of M17.

The synthetic spectra were obtained with the {\sc chianti} database
\citep{dere97,landi06}.  We calculate the X-ray emission coefficient
$j_{\nu}(n,T)$ in the energy range of 0.1-2.4 keV to coincide with
that of ROSAT.  Based on observational works
\citep{dunne03,b1,garciar07} a solar abundance was assumed for
M17\footnote{In \S 4 section we discuss the effect of the metalicity
  on the total X-ray luminosity.}, together with the ionisation
equilibrium model of \citet{mazzotta98}. The $j_{\nu}(n,T)$
coefficients are calculated in the limit of low density 
regime, which results in $j_{\nu}(n,T)\propto n^2$.  To produce X-ray
maps the absorption is included, assuming a uniform hydrogen column
density of $N_{H}=3.2\times 10^{21}\ \mathrm{cm}^{-2}$.

\section{ROSAT image of M17}

M17 was observed with the ROSAT satellite, with the Position
Sensitive Proportional Counter (PSPC) detector. This instrument is
sensitive to X-ray photons with energies in the range $0.1-2.4$ keV
and has an spectral resolution energy of $\sim 40\%$ at 1 keV, with a
field of view of $\sim 2^{o}$, covering the entire dimensions of
superbubble M17 (with an angular size of $\sim 23 \arcmin \times 20
\arcmin$).

We have used an image (binned to $4\arcsec$ per pixel) from
archival ROSAT Catalogue of PSPC WGA sources, which corresponds to the PSPC
observations of M17 (id:WP500311) obtained on September
12-13, 1993, with an exposure time of $6.7$ ks. In order to study the spatial
distribution of the X-ray emission from M17, we have increased the signal
to noise ratio using the standard procedure of smoothing the
image (Dunne et al. 2003, Townsley et al. 2003). To do this, we
convolved the observations with a Gaussian function with an effective
PSF of $5.4$ pixels.
In Figure \ref{f4} we show the original (raw data) image and the
smoothed version.

\section{Results and discussion}

\begin{figure}
 \includegraphics[angle=0,width=0.8\columnwidth]{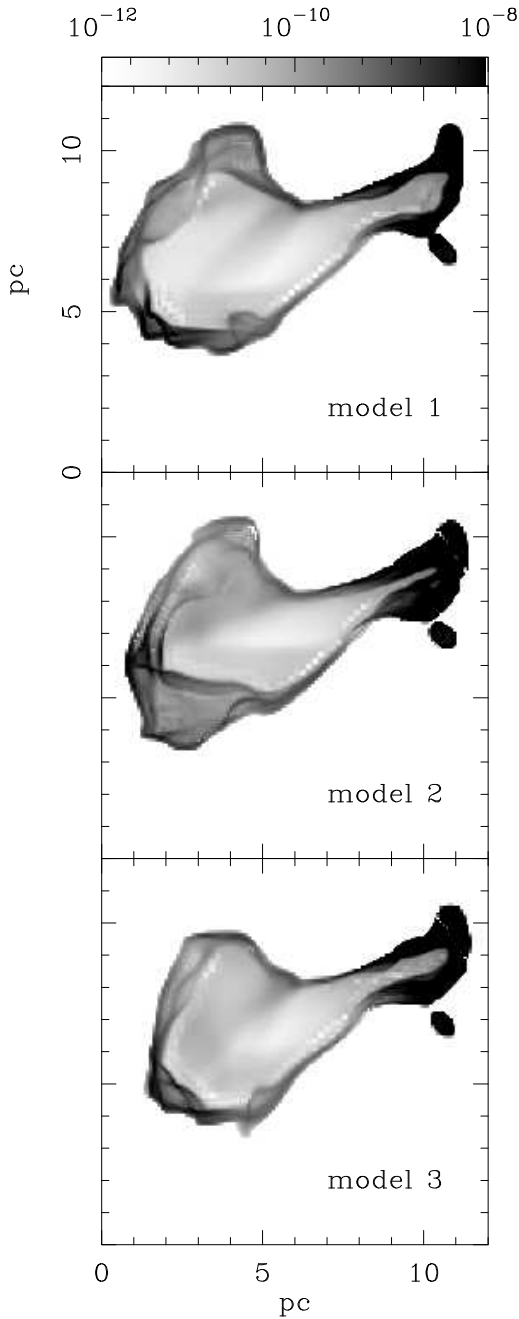}
\caption{Simulated X-ray emission maps obtained from models 1, 2, and
3 (from the top to the bottom) in the energy range $0.1-2.4$ keV, an
absorption due to $N_{H}=3.2\times 10^{21}\mathrm{cm^{-2}}$ has been
included.  Maps corresponding to models 1 and 2 show more evolved
superbubbles (produced by the cluster wind). The bar on the top gives
the logarithmic grey-scale in units of erg cm$^{-2}$ s$^{-1}$
sr$^{-1}$. The vertical and horizontal axis are given in units of pc.}
\label{f3}
\end{figure}

We ran the models up to an integration time of $1.3\times 10^{5}$~yr,
which is the estimated superbubble age (Weaver et al. 1977). In
Figure~\ref{f3} we present X-ray emission maps for the three models
at their latest integration time.  These maps show the absorbed X-ray
emission, assuming a column density of $3.2\times 10^{21}\
\mathrm{cm}^{-2}$ as found from ROSAT observations (Dunne et al. 2003).

The interaction of many individual stellar winds will coalesce into a
common cluster wind, as studied for massive stellar clusters by
\citet{canto00} and \citet{ary07}.
However, unlike those models that consider a uniform ISM, we can see
from Figure~\ref{f3} that the resulting X-ray emission is not
homogeneous. All models show a strong
concentration of X-ray emission in regions to the top right of each
map, especially for model 2.  This enhancement corresponds to strong
shock waves produced by the collision of the cluster wind with the
dense molecular cloud or bars. These bars are flat and 
focus the cluster wind flow
and produce the elongated morphology observed in X-rays.  To the
opposite side the cluster wind can expand more freely producing
through a `champagne'-type flow a superbubble of $\sim 12~\mathrm{pc}$
in size. 
From Figure~\ref{f3}, the structure seen in our simulations at the 
position $x\sim 10\mathrm{pc}$,  $y\sim 7\mathrm{pc}$ corresponds to 
the wind of an isolated star embedded in the the southern bar,
which gets confined due to the high density of its surrounding medium
(see Fig. \ref{f1}).

The surrounding dense bars have another important effect, namely to increase
the total X-ray luminosity $\mathrm{L_X}$. 
This effect depend on the geometry of the bars, which in our models we
have considered to be flat (independent of $z$). If they would have a
concave section, the cluster wind would be more strongly focused,
increasing the  $\mathrm{L_X}$, and the opposite would be true if they
were convex.
To analyse influence of the bars, we have run an additional model (not
presented here), removing them from the numerical setup. In that case
the cluster wind is free to sweep the less dense ISM, and results in a
$\mathrm{L_X}$ 2 orders of magnitude lower than the obtained one from
Models 1, 2, and 3.
This shows the importance of considering the characteristics of the
environment of M17, in particular the dense molecular cloud that surrounds the
open cluster NGC 6618.

In Table \ref{tablum} we list the $\mathrm{L_X}$ obtained for all the models.
We must notice that the X-ray luminosity listed there has been calculated
directly from the X-ray emission coefficients derived from the simulations,
thus no correction for absorption is needed. In other words, it should be
compared with the X-ray luminosity of observations after these observations
have been corrected for absorption.  Models 1, 2, and 3 produce similar
luminosities, all of them on the order of 60 per cent of the observed value
(although Model~1 gives a slightly lower value).  One has to bear in mind that
the gas in the bar is not uniform in reality, but it has high density clumps
(Townsley et al. 2003). The inclusion of such clumps could somewhat increase
the total X-ray luminosity compared with that obtained in our models.

Other factor that can change the obtained $\mathrm{L_X}$ is the
metal abundance. To estimate importance of this factor, we generated new
X-ray coefficients (with the CHIANTI atomic database) considering
different values of the metalicity  $Z$ around of the solar
value. The resulting luminosities are listed in table
\ref{tablum2}. For $Z=0.3$, the X-ray 
luminosity is  2.7 times lower than the obtained for $Z=1$ (solar
abundance). We must note, however, that this is only a rough
estimate. A detailed study of the effect of metalicity on the
X-ray luminosity would have to include the dynamical effect that
results from a modified radiative cooling.
In any case, recent work \citep{garciar07} show that abundances close
to the solar ones are adequate for M17.

Dunne et al.~(2003) obtained a total X-ray luminosity of $5\times
10^{35} \mathrm{erg\ s^{-1}}$ for M17, employing both the Weaver et
al. (1977) bubble model, and the Chu et al.~(1995) superbubble model.
Such value is two orders of magnitude higher than the observed one.
These models are based on the evolution of bubbles (and superbubbles)
into an homogeneous medium, and they are adiabatic, although they
include thermal conduction effects.  In our case, it was not necessary
to include the last physical process because we have obtained a $\mathrm{L_X}$
already compatible with observations, taking only into account the
inhomogeneity of surrounding ISM, and including the radiative cooling.

Although thermal conduction effects should be present, as firstly
introduced in the wind-blown bubble model by Weaver et al. (1977),
there are important discrepancies between the X-ray luminosities
predicted by that model and what is observed. For instance, X-ray
emission should be detected in almost all the known
wind-blown bubbles and superbubbles, but this is not the case (see
Mac Low 2000, for a discussion). In particular, for M17, a direct
application of  Weaver's et al. model (which considers thermal
conduction as a crucial ingredient for the X-ray emission of the hot
interior) predicts an X-ray luminosity
about 100 times higher than what is measured (Dunne et al. 2003).
This leads us to think that perhaps some mechanisms are present that
inhibit thermal conduction, at least in the form that it is treated in
Weaver's et al. model. The presence of magnetic fields is often
invoked as one of the possible inhibiting factors.  In our particular
case, we show that whatever mechanism might be at work, thermal
conduction do not seem to play an important role in the prediction of
X-ray emission from the M17 superbubble.

\begin{table}
\centering
 \begin{minipage}{84mm}
 \caption{Total X-ray luminosities.}
  \begin{tabular}{lc}
  \hline
& $\mathrm{L_X}$ ($10^{33}~\mathrm{erg~s^{-1}}$) \\
ROSAT observation\footnote{Dunne et al. 2003}
&$2.5 $ \\
Model 1 & $1.4 $ \\
Model 2 & $1.6 $ \\
Model 3 & $1.6 $ \\ 
 \hline
\label{tablum}
\end{tabular}
\end{minipage}
\end{table}

%

\begin{table}
\centering
 \begin{minipage}{100mm}
 \caption{Metalicity vs Total X-ray luminosities.}
  \begin{tabular}{lc}
  \hline
$Z$ &$\mathrm{L_X}$ ($10^{33}~\mathrm{erg~s^{-1}}$)\\
0.3&0.6\\
0.5&0.9 \\
0.7 & 1.2 \\
0.9 & 1.4 \\
1.0 &1.6 \\
1.1&1.7\\
1.3&2.0\\ 
 \hline
\label{tablum2}
\end{tabular}
\end{minipage}
\end{table}

\begin{figure}
 \includegraphics[angle=0,width=\columnwidth]{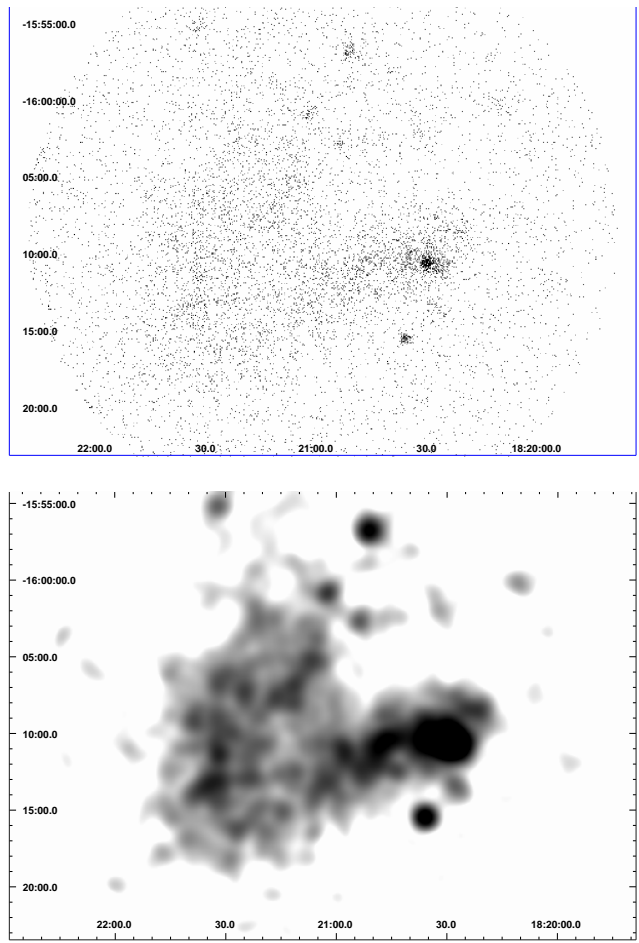}
\caption{ROSAT image of superbubble M17, obtained from archival
data (ROSAT catalogue of PSPC WGA sources) in the energy range
$0.1-2.4$ keV.
{\it Top} panel: raw data binned to $4\arcsec$ per pixel. {\it Bottom} panel:
Smoothed image obtained after convolving with a Gaussian function with
a PSF of $5.4$ pixels.}
\label{f4}
\end{figure}

\begin{figure}
\includegraphics[angle=0,width=\columnwidth]{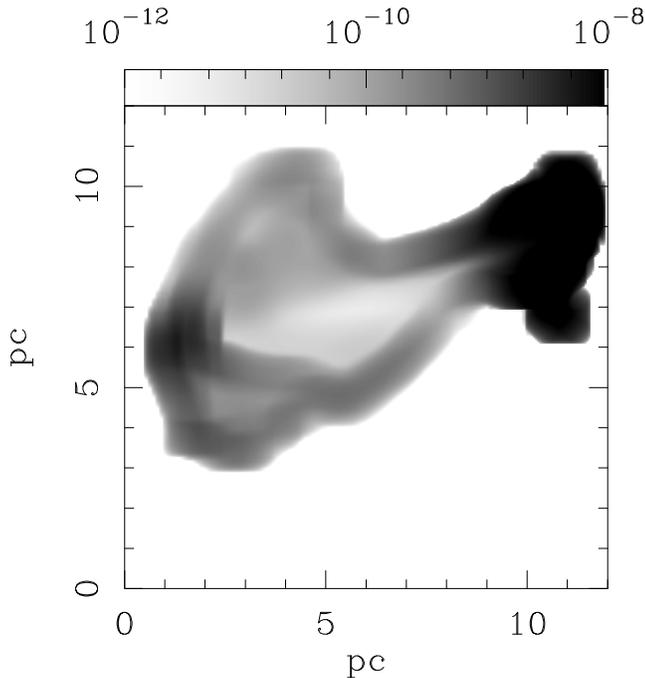}
\caption{Synthetic X-ray emission map of model 2 in the energy range
$0.1-2.4$ keV, smoothed with the same resolution of the ROSAT
observation (Figure~\ref{f4}). The bar on the top is the logarithmic
grey-scale, which is given in units of erg cm$^{-2}$ s$^{-1}$
sr$^{-1}$. Vertical and horizontal axis are given in units of pc.}
\label{f6}
\end{figure}

The X-ray emission distribution for all models (see Figure~\ref{f3})
show a general gradient, with lower emission to the left, in
coincidence with the observations (see Figure~\ref{f4}).  Also in our
models, on the left of the X-ray emitting material, strong emission is
observed in form of filaments (see Figure~\ref{f3}).  These filaments
trace the shock front of champagne flow described above, propagating
into the surrounding ISM. In ROSAT observations, bright regions or
`clumps' can be seen to the East (corresponding to the left, in our
simulations).  These clumps can be produced by the propagation of
strong shock waves into an inhomogeneous medium. However, these features are smoother than those
obtained from the simulations probably due to the difference between
observed and simulated resolutions.  To make a fair comparison we show
in Figure~\ref{f6} a smoothed synthetic map of model 2, produced by
convolving the numerical results with a Gaussian beam in order to
reproduce the same resolution of the ROSAT (smoothed) image in
Fig. \ref{f4}.

\section{Conclusions}

We carried out 3D numerical simulations using the adaptive grid code
{\sc yguaz\'u-a} to model the M17 superbubble. Radiative losses have
been included in the simulations.  Three different runs of the same
model were made using the known positions of the dominant wind sources
on the plane of the sky (corresponding to the $xy$ positions in our
Cartesian grid), and considering three different distributions
along the line of sight (aligned with our $z-$axis).

Our results show that the inclusion of the main features of the
surrounding ISM (i.e. the presence of two dense bars or clouds around
the M17 stars) plays a crucial role to explain both the observed
morphology and the total X-ray luminosity of this object.

On the one hand, the bars produces in all models a `champagne' flow
effect, in which the resulting stellar cluster wind is focused by such
bars, producing an elongated shape in the $x-$~direction, thus reproducing
the elongated shape in the East-West direction of the observations.

On the other hand, the collision of the cluster wind with the bars
maintains a high-temperature, high-density distribution in the region
between the bars. This is a product of the multiple reflected shocks,
and enhances the total X-ray luminosity ($\mathrm{L_X}$).  We obtain
values of the total X-ray luminosity of $\simeq 1.6\times 10^{33}
\mathrm{erg\ s^{-1}}$, which are in agreement with observations.

The different $z-$ stellar distributions employed in our runs had a
rather small effect on both the morphology and $\mathrm{L_X}$. This
could be somewhat expected, because the same mechanical luminosity
input is used for all the runs, and because away of the cluster centre
the common wind that forms should be similar.  Finally, it is
noticeable that we obtain a reasonable estimate of the total X-ray
luminosity without the need of including thermal conduction effects.

\section*{Acknowledgements}
The authors acknowledge support from CONACyT grant 46828-F and 40095-F, and
DGAPA-UNAM grants IN108207, IN100606, and IN117708.  The work of ARG, AE, and
PFV was supported by the `Macroproyecto de Tecnolog\'\i as para la Universidad
de la Informaci\'on y la Computaci\'on' (Secretar\'\i a de Desarrollo
Institucional de la UNAM).  We also would like to thank the computational team
of ICN: Enrique Palacios and Antonio Ram\'\i rez, for maintaining and
supporting our Linux servers, and Mart\'\i n Cruz for the assistance provided.
CHIANTI is a collaborative project involving the NRL (USA), RAL (UK), and the
following Univerisities: College London (UK), Cambridge (UK), George Mason
(USA), and Florence (Italy).

\bsp

\label{lastpage}

\end{document}